\documentclass{amsart}
\newenvironment{nouppercase}{%
  \renewcommand{\uppercasenonmath}[1]{}}{}
\usepackage{enumerate,mathrsfs}
\usepackage{amssymb}
\usepackage{amsmath}
\usepackage{amsfonts}
\usepackage{braket}
\usepackage{color}
\usepackage{graphicx}
\usepackage{tikz}
\usetikzlibrary{calc, shapes, positioning,backgrounds,arrows,automata,backgrounds,fit,decorations.pathreplacing}
\renewcommand{\subset}{\subseteq}

\DeclareMathOperator{\Aut}{Aut}
\DeclareMathOperator{\mult}{mult}
\DeclareMathOperator{\rank}{rank}
\DeclareMathOperator{\spn}{span}

\theoremstyle{definition}
\newtheorem{defn}{Definition}
\newtheorem{example}{Example}
\theoremstyle{theorem}
\newtheorem{theorem}{Theorem}
\newtheorem{lemma}{Lemma}

\title{\text{Quantum structures from association schemes}}
\author{Radhakrishnan Balu$^{\dag}$}
\address{$^{\dag}\text{Computer and Information Sciences Directorate, Army Research Laboratory,
Adelphi, MD, 21005-5069, USA}$.}
\email{radhakrishnan.balu.civ@mail.mil}
\address{$^{\dag}$Department of Mathematics, University of Maryland, College Park, MD 20742}
\email{rbalu@math.umd.edu}
\begin{document}
\begin{nouppercase}
\maketitle
\end{nouppercase}
\begin{abstract}
Starting from an association scheme induced by a finite group and the corresponding Bose-Mesner algebra we construct quantum Markov chains (QMC), their entangled versions, and interacting Fock spaces (IFS) using the quantum probabilistic approach. Our constructions are based on the intersection numbers and their duals Krien parameters of the schemes with examples focused on regular (distance-regular and distance-transitive) graphs.
\end{abstract}
\section {Introduction}
Physical systems often possess  symmetries either at their kinematics, dynamics, or both. Analysis based on the underlying groups provide important insights in ways systems function and also simplify computations of observables and other statistics. We consider association schemes induced by finite groups and so the elements of the algebra are matrices with entries in \{0, 1\} that can be thought of as the adjacency matrix of a graph.  Biane set up Markov chains using ensembles generated from $SU_2$ operations \cite {Biane1989} that resulted in a walk on the dual space. Parthasarathy \cite{KP1990} generalized this construction to compact topological groups by exploiting the Peter-Weyl theorem to fashion the walks on their duals (Plancherel decompositions), the irreducible unitary representations of the groups indexed by the characters of the representations, and constructed the chains. A familiar example of such an evolution is the quantum walk with coins drawn from the compact group $SU_2$ but without the walker degree of freedom which is a coin ensemble. This process, which is a quantum Brownian motion in vacuum state, can be interpreted as a quantum white noise to describe second quantized fields in quantum optics \cite {Accardi2002}. Inspired by these constructions we set up quantum Markov chains on the corresponding hypergroups (duals) of the association schemes which when restricted to a commutative sub-algebra describe classical Markov chains. Entanglement is an important resource in quantum information processing \cite {MUNROE} and processes that generate quantum correlations among the nodes as the evolutions proceed are of interest in local networks. In the next step we provide prescriptions for constructing entangled versions of the chains which embed the classical chains canonically \cite {Accardi2004}. Finally, we identify an interacting Fock space (IFS) that is a multi-dimensional space of polynomials whose indeterminates are members of the association scheme to describe processes on growing graphs. Motivations for our abstract treatment of the quantum structures that arise with association schemes are founded on the quantum walks framework on graphs that are powerful tools in the simulation of complex quantum systems.  Processes on graphs are increasingly more realistic With the steady progress in networking of quantum systems.
\
\section {Association Schemes}
\begin{defn}
Let $X$ be a (finite) vertex set, and let $\mathfrak{X} = \{A_j\}_{j=0}^d$ be a collection of $X\times X$ matrices with entries in $\{0,1\}$. We say that $\mathfrak{X}$ is an \emph{association scheme} if the following hold:
\begin{enumerate}[(1)]
\item $A_0 = I$, the identity matrix;
\item $\sum_{j=0}^d A_j = J$, the all-ones matrix (In other words, the $1$'s in the $A_j$'s partition $X\times X$);
\item For each $j$, $A_j^T \in \mathfrak{X}$; and
\item For each $i,j$, $A_i A_j \in \spn\mathfrak{X}$.
\end{enumerate}
A \emph{commutative} association scheme also satisfies
\begin{enumerate}[(1)]
\setcounter{enumi}{4}
\item For each $i,j$, $A_i A_j = A_j A_i$.
\end{enumerate}
\end{defn}
The above association scheme may be viewed as the adjacency matrices of graphs with a common set of $|\mathfrak{X}| = d$ vertices. Alternately, the scheme can represent 1-distance, 2-distance, ..., d-distance matrices of the same graph. We will take the former view while discussing multi-modal interacting Fock spaces later.
\begin{example} \label{ex:1}
Let $X=G$ be a finite group. For each $x\in G$, let $A_x$ b the the matrix for left translation by $x$ in $\ell^2(G)$. In other words, $A_x$ is the $G\times G$ matrix with
\[ (A_x)_{y,z} = \begin{cases}
1, & \text{if }y=xz \\
0, & \text{otherwise}
\end{cases} \]
for $y,z \in G$. When $e\in G$ is the identity element, we have $A_e = I$, $\sum_{x\in G} A_x = J$, $A_x^T = A_{x^{-1}}$, and $A_x A_y = A_{xy}$. Thus, $\mathfrak{X}:= \{A_x\}_{x\in G}$ is an association scheme.
\end{example}

\begin{example}
\item Let $G$ be a finite group acting transitively on a finite set $X$. Then $G$ also acts on $X\times X$ through the action $g\cdot (x,y) = (g\cdot x, g\cdot y)$ for $g\in G$ and $x,y \in X$. Let $R_0,\dotsc,R_d \subset X\times X$ be the orbits for this action, numbered so that $R_0 = \{ (x,x) : x \in X\}$. (This is an orbit since $G$ acts transitively on $X$.) For each $j=0,\dotsc,d$, let $A_j$ be the $X \times X$ matrix with
\[ (A_j)_{x,y} = \begin{cases}
1, & \text{if } (x,y) \in R_j \\
0, & \text{otherwise.}
\end{cases} \]
Then, one can show, $\mathfrak{X} = \{A_j\}_{j=0}^d$ is an association scheme. It will be commutative if and only if the action of $G$ on $X$ is \emph{multiplicity free}. In other words, the permutation representation of $G$ associated with its action on $X$ decomposes as a direct su, of irreducibles, with no irreducible repeated up to unitary equivalence.
\end{example}

\begin{example} \label{ex:3}
Let $X=G$ be a finite group, and let $K \subset \Aut(G)$ be a group of automorphisms of $G$. Let $\{e\} = C_0,\dotsc, C_d$ be the orbits of $K$ acting on $G$. If $\{A_x\}_{x\in G}$ are as in Example~\ref{ex:1}, define $B_0,\dotsc,B_d$ by
\[ B_j := \sum_{x\in C_j} A_x. \]
Then $\mathfrak{X}:=\{B_j\}_{j=0}^d$ is an association scheme. We call this a \emph{subscheme} of $\{A_x\}_{x\in G}$.

When $K$ is the group of inner automorphisms of $G$ (i.e.\ conjugations by elements of $G$), the orbits $C_0,\dotsc,C_d$ are precisely the conjugacy classes of $G$. Then $\mathscr{B}:=\spn\{B_0,\dotsc,B_d\}$ is the center of the group von Neumann algebra $\mathscr{A}:=\spn\{A_x : x \in G\}$.
\end{example}

\begin {example} The Johnson scheme J (v,k). The vertex set of this scheme is the set of all k-subsets of a fixed set of v elements. Two vertices $\alpha$ and $\beta$ are i-related if $\|\alpha \cap \beta\| = k - i$. This scheme has k classes.
\end {example}

\begin {example} \label {ex: Grassmann}
The Grassmann scheme $J_q (v, d)$. The vertex set is the set of all subspaces of dimension d of the vector space of dimension n over GF(q) (finite field with q elements). Subspaces $\alpha$ and $\beta$ are i-related if $dim(\alpha \cap \beta) = i$. This q-deformed Johnson scheme has d classes, may be thought of as a discrete version of a Grassmannian manifold, and the graph it generates is distance transitive and the basis for our construction of an IFS.
\end {example}
\begin{defn}
The \emph{adjacency algebra} of an association scheme $\{A_j\}_{j=0}^d$ is $\mathscr{A}:=\spn\{A_j\}_{j=0}^d$. Sometimes this is also called the \emph{Bose-Mesner algebra}. It's a unital $*$-algebra of matrices, i.e.\ a von Neumann algebra. It is also closed under the Hadamard (Schur) product.
\end{defn}

Suppose for the moment that our association scheme $\{A_j\}_{j=0}^d$ is commutative. Then, by the spectral theorem, the matrices $A_0,\dotsc,A_d$ are simultaneously diagonalizable. Put differently, the adjacency algebra $\mathscr{A}$ has an alternative basis $E_0,\dotsc,E_d$ of projections onto the maximal common eigenspaces of $A_0,\dotsc,A_d$. Since $\mathscr{A}$ is closed under the Hadamard product, there are coefficients $q_{i,j}^k$ such that
\[ E_i \circ E_j = \frac{1}{|X|} \sum_{k=0}^d q_{i,j}^k E_k \qquad (0 \leq i,j \leq d). \]
The coefficients $q_{i,j}^k$ are called the \emph{Krein parameters} of the association scheme. This leads to a commutative hypergroup. Let $m_j = \rank E_j$, and define $e_j = m_j^{-1} E_j$. Then
\[ e_i \circ e_j = \frac{1}{|X|}\sum_{k=0}^d \left( \frac{m_k}{m_i m_j} q_{i,j}^k \right) e_k. \]

Now, we introduce the dual notion to Krein parameters, the Intersection numbers $p^k_{ij}$ in terms of matrix product
$A_i \bullet A_j= \sum_{k} p^k_{ij}A_k$.
Intuition: In a distance-regular graph (ex: complete graphs, cycles, and odd graphs) the number of paths between a pair of k-distant vertices via i-distant plus j-distant paths is independent of the pair.
We see that association schemes are generalization of groups and hypergroups generalize schemes. One can canonically attach a projective geometries, that are building blocks to quantum systems, to hypergroups \cite {Connes2011}  provide us the motivation for studying these structures in-depth. Projective spaces are difficult to picture in higher dimensions, Bloch sphere is complex projective line, and quantum graph may provide the required to intuition to study them.
\begin{theorem}
For each $i,j$, the mapping $k\mapsto \frac{m_k}{m_i m_j} \frac{q_{i,j}^k}{|X|}$ defines a probability distribution $\mu$ on $\{0,\dotsc,d\}$. If we define
\[ (e_i * e_j)(k) = \frac{m_k}{m_i m_j} \frac{q_{i,j}^k}{|X|}, \]
so that
\[ e_i \circ e_j = \sum_{k=0}^d ( e_i * e_j)(k)\cdot e_k, \]
then $\{e_0,\dotsc,e_d\}$ has the structure of a commutative hypergroup with identity element $e_0 = \frac{1}{|X|}J$ (the all-ones matrix, scaled by $|X|^{-1}$) and involution given by entry-wise complex conjugation.
\end{theorem}
A probability measure is characterized by m-moments $\forall m \ge 1$ and in the context of graphs (Example \ref {ex: Bern}) they correspond to m-step walks from starting and ending at the same vertex.
\begin{example}
Consider the example of the conjugacy classes in a group, as in the last paragraph of Example~\ref{ex:3}. The spectral basis $E_0,\dotsc,E_d$ is essentially given by the irreducible characters $\chi_0,\dotsc,\chi_d$ of the group:
\[ E_j = \frac{\dim(\chi_j)}{|G|} \sum_{x\in G} \chi_j(x)\cdot A_x, \]
with multiplicity $m_j = \dim(\chi_j)^2$. Thus,
\[ e_j = \left[ \dim(\chi_j) \right]^{-1} \cdot \frac{1}{|G|} \sum_{x\in G} \chi_j(x)\cdot A_x, \]
with scaling exactly as occurred in the Parthasarathy's paper \cite{KP1990}. The hypergroup structure that comes from the association scheme perspective is the same as the usual hypergroup structure on $\hat{G}$, namely
\[ (e_i * e_j)(k) = \frac{ \dim(\chi_k) }{\dim(\chi_i) \dim(\chi_j)}\cdot \mult(\chi_k, \chi_i \otimes \chi_k). \]
\end{example}

In terms of association schemes, Parthasarathy's paper \cite{KP1990} that inspired our program is making a quantum Markov chain on the group von Neumann algebra whose transition operator is just Hadamard multiplication by a fixed $e_i$. When this is restricted to the center of the group algebra, we get a classical Markov chain which is just a random walk on the hypergroup $\hat{G}$. We should be able to replicate this for other kinds of association schemes. All we need are two ingredients:
\begin{enumerate}[(1)]
\item A noncommutative association scheme $\mathfrak{X} = \{A_j\}_{j=0}^d$ (for instance, coming from a nonabelian group, or a transitive permutation action that is not multiplicity free).
\item A commutative subscheme of $\mathfrak{X}$. In other words, we need a partition $\{0,\dotsc,d\} = \bigsqcup_{i=0}^{d'} C_i$ such that the matrices
\[ B_i:= \sum_{j\in C_i} A_j \qquad (0 \leq i \leq d') \]
form a \emph{commutative} association scheme.
\end{enumerate}

Then the big adjacency algebra $\mathscr{A} := \spn\{A_0,\dotsc,A_d\}$ is a von Neumann algebra, and the little adjacency algebra $\mathscr{B}:=\spn\{B_0,\dotsc,B_{d'}\}$ is a commutative $*$-subalgebra of $\mathscr{A}$. Let $E_0,\dotsc,E_{d'}$ be the spectral basis of $\mathscr{B}$, with ranks $m_0,\dotsc,m_{d'}$, and let $e_j = m_j^{-1} E_j$ for $0 \leq j \leq d'$, as before.

Fix any $e_i$, and define $T\colon \mathscr{A} \to \mathscr{A}$ by $T(M) = e_i \circ M$ (Hadamard multiplication) for any $M \in \mathscr{A}$. Since $e_i$ is positive, $T$ is completely positive \cite[Theorem~3.7]{Paulsen2002}. Moreover, $\mathscr{B}$ is an invariant subspace of $T$, and $\left. T\right|_{\mathscr{B}}$ describes a classical Markov chain on the state space $e_0,\dotsc,e_{d'}$, corresponding to a random walk on the hypergroup $\{e_0,\dotsc,e_{d'}\}$.

More generally, we could replace $e_i$ with any convex combination of $e_0,\dotsc,e_d$ and obtain the same thing. We have a family of Quantum Markov Chains indexed by $\{0\leq{i}\leq{d}\}$. Later, we will see these transition probability amplitudes will define raising and lowering operators of an interacting Fock space (IFS) \cite {Obata2007} stratified by the conjugacy classes.
\
\section {Entangled Quantum Markov Chains}

Correlations are fundamental quantities in quantum physics from which other relations such as the canonical commutation relations can be derived. Canonical commutation relations in the cases of Bosonic and Fermionic systems follow from Gaussian statistics but the converse is true only in the case of Fock space \cite {Accardi2002}. It is rather easy to generate correlations on a spin chain \cite {RadB2016} by simply applying unitaries chosen randomly on one and two-body (nearest neighbor) terms of the chain. Given a quantum Markov chain a unique set of correlators can be defined using expectation $\mathscr{E}$ at the state $\phi_0$ as:
\begin {equation}
\phi_0(f_0.\mathscr{E}_1(f_1.\mathscr{E}_2(f_2\dots{\mathscr{E}_n({f_n})}\dots))).
\end {equation}
In the above equation the operation "." is usually matrix multiplication giving rise to correlation between observables and if it is replaced by Schur $\circ$ multiplication (Hadamard product) it will lead to entanglement.
\begin {equation}
\phi_0(f_0\otimes\mathscr{E}_1(f_1\otimes\mathscr{E}_2(f_2\dots{\mathscr{E}_n(\mathbb{I}\otimes{f_n})}\dots))).
\end {equation}
Let us formally define a quantum probability space that is used in the following sections.
\begin{defn} A finite dimensional quantum probability (QP) space is a tuple $(\mathscr{H},\mathscr{A}, \phi)$ where  $\mathscr{H}$ is a separable complex Hilbert space, $\mathscr{A}$  is a C* algebra that constitute the event space of orthogonal projections, and $\phi$ is a trace class operator, specifically a density matrix in finite dimensional case, denoting the quantum state. Alternately, we can start with a von Neumann algebra $\mathscr{A}$ and and a state $\phi$ which is a positive linear functional and by GNS construction we can have a separable Hilbert space.\end{defn}

In this section we build entangled Markov chains that are based on the classical chains constructed earlier on the hypergroup $e_0,\dotsc,e_{d'}$ given by the transition operator: $T\colon \mathscr{A} \to \mathscr{A}$ by $T(M) = {e_i}\circ{M}, \forall{M}\in\mathscr{A}$. A family of entangled QMCs can be constructed corresponding to each $e_i$. As we will see later each of these QMCs will contribute to the ladder operators of the interacting fock spaces.

Markov chains are simple yet very powerful tools with wide range of applications in classical probability theory \cite {Motwani1995}. Quantized version of the chains \cite {Szegedy2004}, \cite {RadLiu2017} have received a lot of attention in the recent past. In this section we will build entangled versions of quantum Markov chains defined earlier based on classical chains, by considering the restrictions to centers of the algebra, as established by Accardi et al \cite {Accardi2004},  to construct such processes.
Let S = {1,2,...,d} be a state space of cardinality $|S| = d < \infty$. We consider a classical Markov chain $(S_n)$ with state space S, initial probability distribution $P = (p_j)$ and transition probability matrix $T = (t_{ij})$. Let us fix the orthonormal basis $\ket{e_i}, i\le{d}$ of $\mathbb{C}^{|S|}$ and a vector $\ket{e_0}$ in this basis. We consider the infinite tensor product Hilbert space defined with respect the stabilizing sequence $(\ket{e_0})_n$:
\begin {equation}
\mathscr{H} = \otimes_{\mathbb{N}}^{\ket{e_0}}\mathbb{C}^{|S|}.
\end {equation}
\begin {equation}
\ket{\Psi}_n = \sum\limits_{j_0, j_1, ..., j_n}\sqrt{p_{j_0}}\prod\limits_{\alpha_0}^{n-1}\sqrt{t_{j_\alpha{j_{\alpha+1}}}}\ket{e_{j_0},e_{j_1},...,e_{j_n}}.
\end {equation}
We denote $M_{|S|}$ the algebra of $d\times{d}$ complex matrices and let $\mathscr{A} = M_{|S|}\otimes{M_{|S|}}\otimes{\dots}=\otimes_{\mathbb{N}}M_{|S|}$ be the $C^{*}$-infinite tensor product of $\mathbb{N}$-copies of $M_{|S|}$. That is, we want to define these processes on the quantum probability space $(\mathscr{H}, \mathscr{A}, \rho=\otimes_\mathbb{N}\ket{e_0}^n)$ that would describe the entire evolution in Heisenberg picture.

\begin {defn} An element $A_{\Lambda}\in\mathscr{A}$ (observable) is called localized in a finite region $\Lambda\subset\mathbb{N}$ if there exists an operator $\bar{A_{\Lambda}}\in\otimes_{\Lambda}M_{|S|}$ such that $A_{\Lambda}=\bar{A_{\Lambda}}\otimes\mathbb{I}_{\Lambda^c}$.
\end {defn}
an important property of the wavefunction $\ket\Psi_n$ is that even though it may not converge in the limit it leads to a state defined on localized observables as the following lemma shows.
\begin {lemma} \cite {Accardi2004} For every local obersrvable $A\in\mathscr{A}_{[0,k]},(k\in\mathbb{N})$ one has
\begin {equation}
\langle\Psi_{k+1}, A\Psi_{k+1}\rangle = \lim_{n\rightarrow\infty}\langle\Psi_n, A\Psi_n\rangle=\phi(A).
\end {equation}
\end {lemma} 
\begin {defn}  A quantum state $\phi$ is a homogeneous quantum Markov chain with an initial state $\phi_0$ over $M_{|S|}$ and transition expectation $\mathscr{E}:M_{|S|}\otimes{M_{|S|}}\rightarrow{M_{|S|}}$ if (the Markov property of future is independent of the past given the present holds): 
\begin {equation}
\phi(A_0\otimes{A_1}\otimes\dots\otimes{A_n}\otimes\mathbb{I}\otimes\mathbb{I}\otimes\dots) = \\
\phi_0[\mathscr{E}(A_0\otimes\dots\mathscr{E}(A_{n-2}\otimes\mathscr{E}(A_{n-1}\otimes\mathscr{E}(A_n\otimes\mathbb{I})))\dots)]
\end {equation}
\end {defn}

\begin {defn} An entangled Markov chain, that has applications in describing ground state Hamiltonian of spin chains \cite {Fannes1992},  is a quantum Markov chain $\phi=(\phi_0,\mathscr{E})=(p_i,t_{ij},\ket{e}_i)$ over the algebra $\mathscr{A}$ where $\phi_0$ is a pure state over $M_{|S|}$, $T=(t_{i,j})$ is a stochastic matrix, ${\ket{e}_i}$ is an orthonormal basis, and the transition expectation $\mathscr{E}(.)=V^*.V$ is given by
\begin {align}
V_n\ket{e_{j_n}} &= \sum\limits_{j_{n+1}\in{S}}\sqrt{t_{j_n{j_{n+1}}}}\ket{e_{j}}\otimes\ket{e_{j_n{n+1}}}. \\
V_n^*\ket{e}_i\bra{e}_j &= \sqrt{t_{j_n{j_{n+1}}}}\ket{e}_i.
\end {align} 
It is easy to verify that $V_n$ is an isometry as $V_n^*V_n = \mathbb{I}$. The operator $V_n$ has the following property:
\begin {equation}
\ket{\Psi}_n = \sum\limits_{j_0, j_1, ..., j_n}\sqrt{p_{j_0}}\prod\limits_{\alpha_0}^{n-1}\sqrt{t_{j_\alpha{j_{\alpha+1}}}}\ket{e_{j_0},e_{j_1},...,e_{j_n}} = \sum\limits_{j_0}\sqrt{p_{j_0}}V_{n-1}\dots{V_0}\ket{e_{j_0}}.
\end {equation}
\end {defn}
A unitary evolution of the above construction can be fashioned similar to the Mark chain based walk investigated by Balu et al., using a combination of reflection and swap operators \cite {RadLiu2017}.
Let us extend the notion Schur multiplication $\circ$ to tensor products as a map m defined below:
\begin {equation}
A\otimes{B} = a_{ik}b_{jl}.
\end {equation}
\begin {equation}
(A\circ{B})_{ij} = a_{ij}b_{ij}.
\end {equation}
\begin {align}
m  &: \mathbb{M}\otimes\mathbb{M}\rightarrow\mathbb{M} \\
m(X)_{(i,j)} &= m(A\otimes{B})_{(i,j)}.\\
& = x_{(ii)(jj)} = a_{ij}b_{ij}.
\end {align}

An operator P is called Schur identity preserving if $\mathbb{E}(P(I)) = I$ where the expectation (diagonal projection) gives back a diagonal matrix with all the off diagonal set to zero and diagonal entries preserved. It is called an entangled  Markov operator if 
\begin {equation}
P(I) \neq{I}. \label {MarkovOperator}
\end {equation}
 A transition expectation is a map that is completely positive and identity preserving given by $\mathscr{E} = id\circ(id\otimes{P})$. When the Markov operator is entangled it is called entangled transition expectation and Markov operators can be constructed from transition matrixes of classical Markov chains as follows:
\begin {equation}
P(A)_{ij} = \sum\limits_{k,l=1}\sqrt{t_{ik}t_{jl}}a_{kl}.
\end {equation}

The operator T, that projects onto a specific adjacency, restricted to the center encodes the information of probability of moving out of a state that leads to a Markov operator. Whereas, the entangled Markov operator P defined by the equation \eqref {MarkovOperator} encodes probability amplitudes superposed. \\
Transition expectation: $\mathbb{E}\colon \mathscr{A}\otimes\mathscr{A} \to \mathscr{A}$ by 
\begin {align*}
\mathbb{E}(M\otimes{N}) &= m \circ [\mathbb{I}\otimes{T}](M\otimes{N}), \forall{M,N}\in\mathscr{A}.\\
&= m \circ [M\otimes{T(N)}].\\
&= m \circ [M\otimes{e_i}\circ{N}]. \text{ \color{blue} QMC with diagonally embedded classical chain.}\\
T &= e_0\circ{M} \text {  is a non-entanglement Markov operator as P(I) = I.} \\
T(M) &= e_i \circ M. \\
\left. T\right|_{\mathscr{B}} &\text{  is has the stochastic matrix  } t^i_{jk}. \\
P(A)_{ij} &= \sum\limits_{k,l=1}\sqrt{t_{ik}t_{jl}}a_{kl}.\\
\hat{\mathbb{E}}^i(M\otimes{N}) &= m \circ [M\otimes{P(N)}]. \text{ \color{blue} An entangled QMC based on P defined above.} \\
\end {align*} 
\begin {theorem}
Family of QMCs $\hat{\mathbb{E}}^i$ indexed by $\{0\leq{i}\leq{d}\}$ are entangled.
\end {theorem}

\section {Interacting Fock spaces}
Interacting Fock spaces, a theory of orthogonal polynomials, are a generalization of the usual symmetric and anti-symmetric Fock spaces that have applications in quantum optics \cite {Accardi2002} and graph theory \cite {Obata2007}. The IFS framework based on quantum probability can be used to describe Bosonic fields with white noise processes used in quantum optics. Unlike in the classical case there are several different stochastic independence that can be formulated in the quantum context that are relevant in the IFS framework. These different stochastic independences can be cast as various graph products based on the monadic operation and correspondingly various central limit theorems manifest for the asymptotics of growing graphs. In a quantum probability space $(\mathscr{A}, \phi)$ the usual commutative independence $(\phi(bab) = \phi(a)\phi(b^2); a,b \in \mathscr{A})$ such as the one assumed in quantum optics leads to conjugate Brownian motions (measured as quadratures) in the limit. The monotone independence $(\phi(bab) = \phi(a)\phi(b)^2; a,b \in \mathscr{A})$ that is relevant in quantum walks leads to arcsin-Brownian motion (double-horn distribution) aymptotically and the other two are free and Boolean independences not focused in this work. In the graph context, the independence notions are defined in terms of products of graphs.
\begin {defn} An IFS associated with the Jacobi sequence $\{\omega_n\}, (\omega_m = 0) \Rightarrow \forall n \ge m, \omega_n = 0, \{\alpha_n\}, \alpha_n \in \mathbb{R}$ is a tuple $(\Gamma \subset \mathscr{H}, \{\Phi_n\}, B^+, B^-, B^\circ)$ where $\{\Phi_n\}$ are orthogonal polynomials and $B^\pm \Phi_n$ spans $\Gamma$. The mutually adjoint operators $B^+, B^-$ and $B^\circ$ satisfy the relations
\begin {align*}
B^+ \Phi_n = \sqrt{\omega_{n+1}} \Phi_{n+1}. \\
B^- \Phi_n = \sqrt{\omega_n} \Phi_{n-1}; B^- \Phi_0 = 0. \\
B^\circ \Phi_n = \phi_n.
\end {align*}
\begin {equation}
xP_n (x) = P_{n+1} (x) + \omega_n P_{n-1} (x) + \alpha_{n+1} P_n (x).
\end {equation}
\end {defn}
With the above IFS we can associate a graph with an adjacency matrix 
$T = \begin{bmatrix} \alpha_1 & \sqrt{\omega_1}  \\
\sqrt{\omega_1} & \alpha_2 & \sqrt{\omega_2}  \\
& \sqrt{\omega_2} & \alpha_3 & \sqrt{\omega_3}  \\
 & & \ddots & \ddots & \ddots & \\
& & & \sqrt{\omega_{n-1}} & \alpha_n & \sqrt{\omega_n} \\
  & & & & \ddots & \ddots & \ddots 
\end {bmatrix}$
that has the quantum decomposition $T = B^+ + B^- + B^\circ$. The sequence $\{\Phi_n\}$ represents fixing a vertex and stratifying (partitioning based on distance from the fixed vertex) the graph with V set of vertices. Let us fix the Hilbert space $\mathscr{H} = l^2(V)$ of the graph. 

\begin{figure}
 \center {
 \begin{tikzpicture}[->,>=stealth',shorten >=1pt,auto,node distance=1.5cm,
  thick,main node/.style={ellipse,fill=green!20,draw,font=\sffamily\Large\bfseries}]
  \node[main node] (1) {\hspace{0.5cm} };
  \node[main node] (2) [right of=1] {\hspace{0.5cm } };
  \node[main node] (3) [below of=1] {0};
  \node[main node] (4) [below of=2] {1};
  \node[main node] (5) [right of=2] {\hspace{0.5cm } };
  \node[main node] (6) [below of=5] {2};
  \node[main node] (7) [right of=5] {\hspace{0.5cm } };
  \node[main node] (8) [below of=7] {3};
  \node[main node] (9) [right of=7] {};
  \node[main node] (10) [below of=9] {};
  \node[main node] (11) [right of=9] {\hspace{0.5cm } };
  \node[main node] (12) [below of=11] {\hspace{0.5cm } };
  \node[main node] (13) [right of=11] {\hspace{0.5cm } };
  \node[main node] (14) [below of=13] {m};
  \node[main node] (15) [right of=13] {\hspace{0.5cm } };
  \node[main node] (16) [below of=15] {\hspace{0.5cm } };

  \path[every node/.style={font=\sffamily\small}]
    (1) edge[-] node {}  (2)    
    (2) edge[-] node {} (4)        
          edge[right] node[right=1mm] {}(6)
    (3) edge[-] node[right=1mm] {}(1)
          edge[left] node[right=1mm] {}(2)        
    (4) edge[-] node[right=1mm] {}(3)
         edge[-] node[right=1mm] {} (6)
    (5) edge[-] node[right=1mm] {} (6)
         edge[-] node[right=1mm] {} (2)
    (7) edge[-] node[right=1mm]{} (8)
         edge[-] node[right=1mm] {}(5)
    (6) edge[-] node[right=1mm] {}(8)
          edge[right] node[right=1mm] {}(7)
    (8) edge[-] node[right=1mm]{} (10)
    (7) edge[-] node[right=1mm]{} (9)
          edge[right] node[right=1mm]{} (12)
    (9) edge[-] node[right=1mm]{} (11)
    (11) edge[-] node[right=1mm]{} (12)
           edge[-] node[right=1mm]{} (13)
    (10) edge[-] node[right=1mm]{} (12)
    (12) edge[-] node[right=1mm]{} (14)
            edge[right] node[right=1mm]{} (13)
    (13) edge[-] node[right=1mm]{} (14)
            edge[right] node[right=1mm]{} (16)
            edge[-] node[right=1mm]{} (11)
    (15) edge[-] node[right=1mm]{} (16)
             edge[-] node[right=1mm]{} (13)
    (14) edge[-] node[right=1mm]{} (16);
    \end{tikzpicture}
  
  \caption{\label{fig:CoinGraph}
     The m-moment of Bernoulli trial represented as a random walk on a graph as an m-step evolution starting and ending at the same node. The directed part may be viewed as a comb product of two graphs each with two nodes and the whole graph may be seen as cartesian product of the two graphs. In the comb product case (monotone stochastic independence), quantum walk evolution, the asymptotic distribution is arcsin-Brownian motion and in the cartesian product case (commutative independence) the long time limit is a non-commutative Brownian motion. 
  }
}
\end{figure}
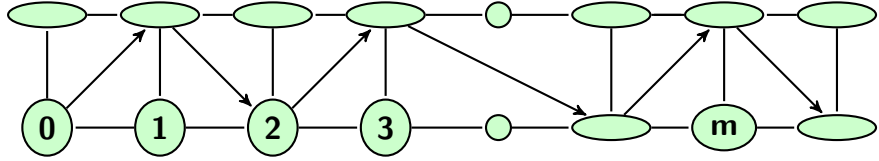

\begin {figure}
\begin{tikzpicture}[->,>=stealth',shorten >=1pt,auto,node distance=2cm,
  thick,main node/.style={circle,fill=blue!20,draw, 
  font=\sffamily\Large\bfseries,minimum size=10mm},
  sec node/.style={draw}]
  \node[main node] (0) {0};
  \node[main node] (1) [right of=0] {1};
  \node[main node] (2) [right of=1] {2};
  \node[sec node] (3) [right of=2] {...};
  \node[main node] (4) [right of=3] {n};
  \node[sec node] (5) [right of=4] {...};
  \path[every node/.style={font=\sffamily\small,
  		fill=white,inner sep=1pt}]
    (0)  edge [loop below] node {$b_0$} (0)
          edge [bend left=60] node[right=1mm] {$c_1$} (1)
    (1)  edge [loop below] node {$b_1$} (1)
          edge [bend left=50] node[left=1mm] {$a_0$} (0)
          edge [bend left=60] node[right=1mm] {$c_2$} (2)
    (2) edge [loop below] node {$b_2$} (2)
          edge [bend left=50] node[left=1mm] {$a_1$} (1)
          edge [bend left=60] node[right=1mm] {$c_2$} (3)
    (3) edge [bend left=50] node[left=1mm] {$a_2$} (2)
         edge [bend left=60] node[right=1mm] {$c_n$} (4)
    (4) edge [loop below] node {$b_n$} (4)
          edge [bend left=50] node[left=1mm] {$a_n$} (3)
          edge [bend left=50] node[right=1mm] {} (5)
    (5) edge [bend left=50] node[left=1mm] {} (4);
 \end{tikzpicture} 
\caption{\label{fig:digraph} Weighted digraph with an adjacency matrix } 
$T = \begin{bmatrix} b_0 & a_0 & & & &  & \\
c_1 & b_1 & a_1 & & &  & \\
 & \ddots & \ddots & \ddots & \\
  & & c_n & b_n & a_n & \\
  & & & \ddots & \ddots & \ddots 
\end {bmatrix}$
\end {figure}
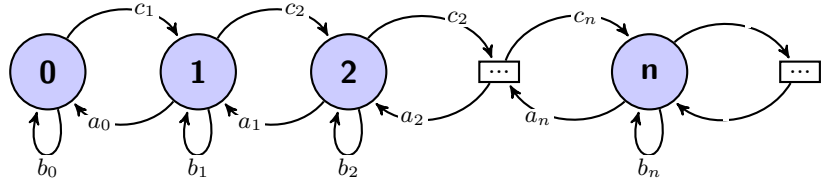

\begin {example} \label {ex: Bern} Bernoulli trial (Figure \ref {fig:CoinGraph}): This process produces an ensemble that is a classical coin toss with the probability measure $\mu = \frac{1}{2}\delta_{-1} + \frac{1}{2}\delta_1$ whose moment sequence is $M^m_\mu = \int_\infty ^\infty x^m \mu(x) = 1$ if m is even and 0 when m is odd. Now, let us consider a graph with two nodes $(e_0, e_1)$ with an edge connecting them. The adjacency matrix for the graph in the standard basis ($\{\begin{bmatrix} 0 \\ 1 \end{bmatrix}, \begin{bmatrix} 1 \\ 0 \end{bmatrix}\}$) of $\mathbb{C}^2$ is $T =  \begin{bmatrix} 0 & 1 \\ 1 & 0 \end{bmatrix}$. In quantum probability we have $\langle e_0, A^m e_0 \rangle = $ if m is even and 0 when m is odd, that is, it reproduces the classical probability measure. In other words, the quantum random variable A reproduces in the vacuum state $e_0$ the moment sequence of the classical coin toss and the fact that it has the canonical decomposition $A = \begin{bmatrix} 0 & 1 \\ 0 & 0 \end{bmatrix} + \begin{bmatrix} 0 & 0 \\ 1 & 0 \end{bmatrix}$ we can say that Bernoulli trial has a quantum decomposition. The classical moment sequence has the new interpretation in the graph context as the number of m-step walks starting from the vertex $e_0$ and ending in it (polynomials of A). The same analysis holds for the state $e_1$ as the vertices of a graph are equivalent any arbitrary vertex can be used to define the vacuum state. Also when a biased coin, produces heads with probability p and tails with probability (1 - p), is used at the quantum state $\phi(A) = p*\alpha e_0 + (1 - p)*\beta e_1$, where $A = \alpha \ket{e_0}\bra{e_0} + \beta \ket{e_1}\bra{e_1}, \alpha^2 + \beta^2 = 1$ the same ensemble is produced. In Hadamard quantum walk on the integer line the observation that the probability amplitudes vanish at odd positions can be understood from the non-existent odd-moments interpretation. So, growing a distance regular graph preserves the moments, of all possible order, information and taking the limit on the size of the graph would provide the asymptotics distribution. 
\end {example}
\begin {example} A weighted directed graph (Figure \ref {fig:digraph}) that is an infinite Markov chain can be represented by an adjacency matrix and an IFS.
\end {example}

\begin {example}
For the Bosonic (symmetric) Fock space we have $\omega_n = n; \alpha_n = 0$.
For the Fermionic (anti-symmetric) Fock space the Jacobi parameters are $\omega_1 = 1; \omega_n =0, n > 1; \alpha_n = 0$.
\end {example}

In this section we construct an IFS for the commutative association scheme with d classes (example \ref {ex: Grassmann}) $J_q (d+k, d)$: $\mathfrak{X}:=\{B_j\}_{j=0}^d$, graph with diameter k, similar to the development of multi-dimensional orthogonal polynomials by Accardi \cite{Accardi2017}. For example, association schemes induced by finite cyclic groups are commutative and as a consequence the schemes are self-duals. In our case the variables of the orthogonal polynomials are matrices and the algebra is also closed under Schur multiplication $\circ$ and thus a *-algebra with the ladder operators (CAPs) of the IFS can be defined in terms of the parameters of the association scheme. These are positive real numbers and can be normalized to become a probability measure with their square root interpreted as probability amplitudes.The classes of the association schemes are referred to as modes and they represent different graphs with common vertices. Orthogonal polynomials in finite number of variables were treated in \cite {Stan2004} and the commutation relations between the ladder operators derived and here our focus is to determine the Jacobi parameters. IFS are a convenient framework to define quantum walks on regular graphs and then to establish their aysmptotics (\cite {Konno2013} \cite{Obata2007}).

\newcommand{\D}{7} 
\newcommand{\U}{7} 

\newdimen\R 
\R=3.5cm 
\newdimen\L 
\L=4cm

\newcommand{\A}{360/\D} 

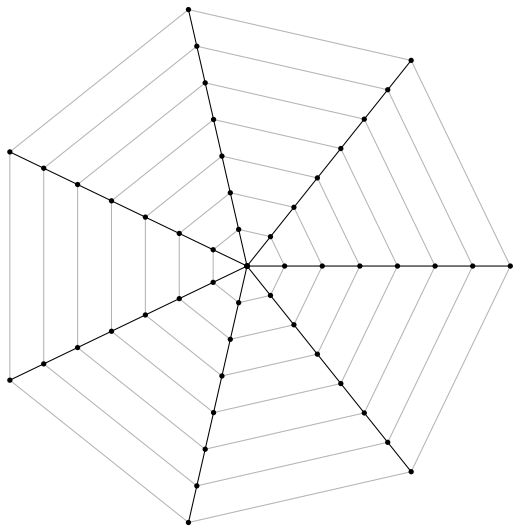
\begin{figure}[htbp]
 \centering

\begin{tikzpicture}[scale=1]
  \path (0:0cm) coordinate (O); 

  \foreach \X in {1,...,\D}{
    \draw (\X*\A:0) -- (\X*\A:\R);
  }

  \foreach \Y in {0,...,\U}{
    \foreach \X in {1,...,\D}{
      \path (\X*\A:\Y*\R/\U) coordinate (D\X-\Y);
      \fill (D\X-\Y) circle (1pt);
    }
    \draw [opacity=0.3] (0:\Y*\R/\U) \foreach \X in {1,...,\D}{
        -- (\X*\A:\Y*\R/\U)
    } -- cycle;
  }

  \path (1*\A:\L) node (L1) {};
  \path (2*\A:\L) node (L2) {};
  \path (3*\A:\L) node (L3) {};
  \path (4*\A:\L) node (L4) {};
  \path (5*\A:\L) node (L5) {};
  \path (6*\A:\L) node (L6) {};
  \path (7*\A:\L) node (L7) {};

\end{tikzpicture}
\caption{Spidernet with Jacobi parameters $\omega_1 = 1; \omega_2 = \omega_3 = \dots = q$ - an example of a stratified graph on which a quantum walk can be fashioned.}
\label{fig:spiderweb}
\end{figure}

Let us denote the *-algebra of matrices on $B_i\in\mathfrak{X}$ as $\mathcal{P} = \mathbb{C}[(B_i),0\leq{n}\leq{d}]$ which are Schur polynomials in the dual representation.  The linear generators of the algebra are the monomials $M = B^{n_1}_1,\dots,B^{n_d}_d$ and $deg(M) = \sum_j (n_j) = n$ with $1_\mathcal{P}$ as as unity satisfying $\langle{1}_\mathcal{P},{1}_\mathcal{P}\rangle = 1$.  We get the symmetric tensor commutative algebra (Bosonic) $\sum_{n\in\mathbb{N}}\mathcal{P}_n^0$ that is graded by the correspondences $e_j\in\mathbb{C}^d\rightarrow{B_j}$ and $\otimes_{sym}(\mathbb{C}^d) = \sum_{n\in\mathbb{N}}\mathcal{P}_n^0 \equiv\mathcal{P}$. Here, $\mathcal{P}_n^0$ is the span of monomials of degree n. We can also build a $\mathbb{Z}_2$-graded *-algebra using anti-symmetric tensor products (Fermionic) as $\otimes_{asym}(\mathbb{C}^d) = \sum_{n\in\mathbb{N}}\mathcal{P}_n^0 \equiv\mathcal{P}^a$. The natural pre-inner product $\langle{B_1},B_2\rangle = tr(B_1^*{B_2}) = sum(B_1\circ{B_2})$ of our (Bose-Mesner in the dual) algebra extends to a pre-inner product on gradations of $\mathcal{P}$ which in turn induces a state $\phi(B_1^*B_2) = tr(B_1^*{B_2})$. Gradation is an algebraic property independent of the measure but when orthogonality is based on the state $\phi$, $\mathcal{P}$ produces orthogonal quantum decomposition of $B_i$s. It is obviously a product state on $\mathcal{P}$ and the filtration is constructed in the usual way. 
\begin {align*}
\mathcal{P}_{n]} &= \text {linear span of monomials of degree n}. \\
P_{n]}:&\mathcal{P}\rightarrow \mathcal{P}_{n]}. \\
P_n &= P_{n]} - P_{n - 1]}.
\end {align*}
We have,
\begin {align*}
a^+_{j\mid{n}} &= P_{n+1} X_j P_n. \\
a^-_{j\mid{n}} &= P_{n-1} X_j P_n. \\
a^+0_{j\mid{n}} &= P_{n} X_j P_n. \\
a^\epsilon_j &= \sum_j a^\epsilon_{j\mid{n}}. \\
B_j &= a^+_j + a^0_j + a^-_j. \\
tr(B_j^*, B_i) &= 0.
\end {align*}

The Jacobi relation in this case is similar to the polynomials in real indeterminates and given by:
\begin {align} \label {eq:Jacobi-3-term}
B_j P_n &= P_{n+1}B_j P_n + P_n B_j P_n + P_{n-1} B_j P_n; 1\leq{j}\leq{d}. \\
B_j P_n &= a^+_{j\mid{n}} + a^0_{j\mid{n}} + a^-_{j\mid{n}}. \\
\end {align}
Our IFS, with generalized Jacobi parameters as positive definite kernels and hermitian matrices, is fashioned on the procedure described in \cite{Accardi2017b} where the n-th level spaces defined recursively in terms of (n-1)-th level spaces. Let us the define the pre-Hilbert space that is a linear span of product vectors of individual modes $\{(\Phi_{1,n}.\Phi_{2,n}\dots\Phi_{d,n}.\Phi_0)\}$ with the vacuum vector:
\begin {align*}
\Phi_0 &= 1_\mathcal{P}. \\
\Phi_{j,0} &= \Phi_0, 0\leq{j}\leq{d}.
\end {align*}

\begin {align*}
\Phi_{j,n} &=  a^+_{jn}\dots{a}^+_{j1}\Phi_0. \\
a^0_{j\mid{0}}\Phi_0 &= p^0_{1j}\Phi_0.\\
a^+_j\Phi_{j,n} &= \sqrt{p^{n+1}_{j;1,n}p^{n}_{j;1,n+1}}.\\
a^-_j\Phi_{j,n} &= \sqrt{p^{n}_{j;1,n-1}p^{n-1}_{j;1,n}}; a^-\Phi_{j,0} = 0.\\
a^o_j\Phi_{j,n} &= p^n_{j;1,n}\Phi_{j;1,n}.\\
\end {align*}
We have d CAPs (creation, annihilation, and preservation operators) one for each mode of the system and they are related to the parameters as follows:
In the case of 1-D polynomials the stratification of the corresponding graph, creation, annihilation, and preservation operators of the algebra can be defined on $\mathscr{A}$ using the natural ordering of the association scheme $0\leq{n}\leq{d}$ as
\begin {align*}
V_n &= \{x\in{C_i}\}.\\
\Phi_n &= \|V_n\|^{-\frac{1}{2}}\sum_{x\in{V_n}}\delta_x.\\
\omega_\epsilon(x) &= p^n_{j;1,n+\epsilon}, \text {  if  } x\in{V_n}. 
\end {align*}
We have the following asymptotic result that is a consequence of the quantum central limit theorem (QCLT) for a growing DRG theorem 6.10 \cite {Obata2007} applied to individual modes.
\begin {theorem}
Let $\mathscr{A}$ be a *-algebra corresponding to the Grassman association scheme $J_q (n, d): \{A_i\}; \{1\leq{i}\leq{d}\}$, with the intersection numbers $\{p^k_{ij}\}$ and $\Gamma(\mathscr{G}) = span(\Phi_d)$. Then, we have the product IFS with the intersection numbers for individual modes given by: 
\begin {align*}
p^{n - 1}_{j;1, n} &= (2 - n)(v - n), &n = 1, \dots, min\{d, v - d\}. \\
p^n_{j; 1, n} &= n(v - 2), &n = 1, \dots, min\{d, v - d\}. \\
p^n_{j; 1, n} &= n(v - 2n). \\
p_{1,1}^0 &= d (n - d).
\end {align*}
\begin {proof} Our Grassmann scheme represents d graphs with a diameter of 1 and the d modes are independent. Thus, in the limit we have a finite product of measures of individual modes whose convergence is guaranteed by QCLT.
\end {proof}
\end {theorem} 
\
\section {Summary and Conclusions}
We constructed quantum Markov chains that evolve on the hypergroups of an association scheme generated by a finite group. We then synthesized entangled versions of the QMCs. We identified an IFS by considering the Bose-Mesner algebra that is closed under matrix multiplication and their CAPs expressed in terms of the intersection numbers. The above analysis can be carried out in the dual space
by assigning an origin for the graph representing each class of the association scheme which will lead to a vacuum state and a product state for the whole scheme. This opens up ways to construct squeezed states and non-classical quantum states such as the Schr$\ddot{o}$dinger CAT states on IFS similar to the Fock space of a quantum harmonic
oscillator. With the construction of new IFS quantum walks evolving on modes and strata degrees of freedom can be constructed. Other possibilities include quantum walk evolution that could be fashioned with multiple walkers, one per mode, and their interactions to model processes on graphs.
\
\section {Acknowledgements}
The author is grateful to and Joseph W.Iverson (jiverson@math.umd.edu) for introducing the facinating topic of association schemes and acknowledges his contribution to the first result. 
\bibliographystyle{abbrv}

\end{document}